\documentclass[aps,prd,showpacs,nofootinbib]{revtex4}
\usepackage{latexsym,bm,amsmath,amssymb,xfrac,xcolor,tensind}
\tensordelimiter{?}
\tensorformat{}
\DeclareMathAlphabet{\mathpzc}{OT1}{pzc}{m}{it}

\def\p{\partial}
\def\m{\mu}
\def\n{\nu}
\def\a{\alpha}

\def\d{\delta}
\def\e{\eta}
\def\eps{\epsilon}
\def\f{\phi}
\def\k{\kappa}

\def\L{\Lambda}
\def\r{\rho}

\def\S{\Sigma}
\def\t{\theta}

\def\vf{\varphi}
\def\nn{\nonumber}
\def\sq{\sqrt}
\def\sqdet{\sq{-g}}
\def\sqh{\sq{h}}
\def\cD{\mathcal{D}}
\def\dif{\mathrm{d}}
\def\goesto{\rightarrow}

\begin{document}
	
		\title{More on the classical double copy in three spacetime dimensions}
		
			\author{Mehmet Kemal G\"{u}m\"{u}\c{s}}
		\email{kemal.gumus@metu.edu.tr}
		
		\affiliation{Department of Physics, Faculty of Arts and Sciences,\\
			Middle East Technical University, 06800, Ankara, Turkey}
		\date{\today}
		
	\author{G{\"o}khan Alka\c{c}}
	\email{gokhanalkac@hacettepe.edu.tr}
	
	\affiliation{Physics Engineering Department, Faculty of Engineering, Hacettepe
		University, 06800, Ankara, Turkey}
	
\begin{abstract}
	
It is well known that general relativity (GR) in three spacetime dimensions (3D) has no well-defined Newtonian limit. Recently, a static solution mimicking the behavior of the expected Newtonian potential has been found by studying the classical double copy of a point charge in gauge theory \cite{CarrilloGonzalez:2019gof}. This is the first example where the vacuum solution in the gauge theory leads to a nonvacuum solution on the gravity side. The resulting energy-momentum tensor was attributed to a free scalar ghost field; however, alternatively, the source can be seen as one resulting from a spacelike perfect fluid.  In this paper, we first give an alternative derivation of the solution where there is no need to perform a generalized gauge transformation to obtain a quadratic Lagrangian without propagating ghost fields. Then, we present a stationary version of the solution and show that the scalar field interpretation of the source does not survive in this case, leaving the spacelike fluid as the only possibility. We give the gauge theory single copy of our solution and comment on the implications of our results on the validity of the classical double copy in 3D. The effect of the cosmological constant is also discussed.

\end{abstract}
	\maketitle	
	
\section{Introduction}

The idea of the double copy, also named BCJ double copy after its discoverers Bern, Carrasco and Johansson, emerged as a relation between the scattering amplitudes in gauge and gravity theories \cite{Bern:2008qj}. A gluon amplitude in the gauge theory is expressed as a sum of cubic graphs which schematically takes the form
\begin{equation}
	A_{\mathrm{Gluon}}=\sum_{i} \frac{n_{i} c_{i}}{d_{i}},
\end{equation}
where $c_i$, $n_i$ and $d_i$ represent the color factors, the kinematic factors and the propagators of each graph respectively. This sum can be written in different ways by making use of generalized gauge transformations, i.e., gauge transformations and field redefinitions. When it is expressed in the so-called color-dual gauge, where the kinematic factors obey the same algebra as the color factors, the graviton amplitude in the gravity theory can be obtained by replacing the color factors $c_i$ by another set of kinematic operators $\tilde{n}_i$ as follows
\begin{equation}
	A_{\mathrm{Graviton}}=\sum_{i} \frac{n_{i} \tilde{n}_{i}}{d_{i}}.
\end{equation}
The new kinematic factors $\tilde{n}_i$ should be also in the color-dual gauge but, in general, they may be taken from a different gauge theory. Therefore, one obtains a gravity = (Yang-Mills)$^2$-type relationship where the gravity theory is called the ``double copy" of the two Yang-Mills (YM) theories, while each YM theory is referred to as a ``single copy" of the gravity theory.

 From the string theory point of view, it is not surprising to obtain such a relation since, at the tree level, it is equivalent to the celebrated  KLT (Kawai, Lewellen, Tye) relations \cite{Kawai:1985xq} between open and closed string amplitudes in the large string tension limit. In addition to the results from the tree level \cite{BjerrumBohr:2009rd,Stieberger:2009hq,Bern:2010yg,BjerrumBohr:2010zs,Feng:2010my,Tye:2010dd,Mafra:2011kj,Monteiro:2011pc,BjerrumBohr:2012mg}, there is accumulating evidence for the double copy to work at loop level 	\cite{Bern:2010ue,Bern:1998ug,Green:1982sw,Bern:1997nh,Carrasco:2011mn,Carrasco:2012ca,Mafra:2012kh,Boels:2013bi,Bjerrum-Bohr:2013iza,Bern:2013yya,Bern:2013qca,Nohle:2013bfa,Bern:2013uka,Naculich:2013xa,Du:2014uua,Mafra:2014gja,Bern:2014sna,Mafra:2015mja,He:2015wgf,Bern:2015ooa,Mogull:2015adi,Chiodaroli:2015rdg,Bern:2017ucb,Johansson:2015oia,Carrasco:2015iwa} and to all orders in certain kinematic limits \cite{Oxburgh:2012zr,White:2011yy,Melville:2013qca,Luna:2016idw,Saotome:2012vy,Johansson:2013nsa,Johansson:2013aca}. Therefore, it is natural to ask whether these results are an indication of a nonperturbative relation between gravity and gauge theories at the perturbative level. The first step to answer this question was taken in \cite{Monteiro:2014cda}, leading to a new research program which extends the double copy from scattering amplitudes to classical solutions. The  idea is to consider stationary metrics of the Kerr-Schild (KS) form
 \begin{equation}
 	g_{\m \n}=\eta_{\mu \nu}+\k\, h_{\mu \nu}, \qquad \partial_0 g_{\mu \nu}=0\label{met}
 \end{equation}
where the deviation $h_{\mu \nu}$ from the background Minkowski space $\eta_{\mu \nu}$ is constructed from a scalar $\phi$ and a vector $k_\mu$ as
\begin{equation}
	h_{\mu \nu}=\f\, k_\mu k_\nu.\label{KS}
\end{equation} 
Here, the vector $k_\mu$ is null and geodesic with respect to both the background metric $\eta_{\mu \nu}$ and the full metric $g_{\mu \nu}$. When the metric is in the KS form (see \cite{Stephani:2003tm} for a review), the Ricci tensor with mixed indices becomes linear in the deviation $h_{\mu \nu}$ and the trace-reversed Einstein equations take the form
\begin{align}
	{R^\m}_{\n}&=\frac{\k}{2}\left(\p^\a \p^\m h_{\n\a} + \p^\a  \p_\n {h^\m}_{\a} -\p^\a\p_\a{h^\m}_\n \right).\nn\\
	&=\frac{\k^2}{2}\left[{T^\m}_\n-\frac{1}{d-2}{\d^\m}_\n T\right],\label{einstein}
\end{align}
where $\k^2=8\pi G$. Choosing $k^0=+1$, the $\m0$ components become
\begin{equation}
\partial_{\nu}\left[\partial^{\mu}\left(\phi k^{\nu}\right)-\partial^{\nu}\left(\phi k^{\mu}\right)\right]=\k\left[?T^{\mu}_0?-\frac{1}{d-2}?\delta^{\mu}_{0}?T\right].\label{muzero} 
\end{equation}
It is easy to see that if one makes the following identifications \cite{Monteiro:2014cda}
\begin{equation}
	A_\m\equiv\phi \,k_\m,\qquad \qquad g\equiv\frac{\k}{2}, 
\end{equation}
 (\ref{muzero}) become the Abelian Yang-Mills equations
\begin{equation}
\partial_{\nu}F^{\nu\mu}=gJ^{\mu},\label{YM}
\end{equation}
where $F_{\m\n}=2\p_{[\m}A_{\n]}$ is the field strength, $g$ is the gauge coupling and the source is given by\footnote{See \cite{Carrillo-Gonzalez:2017iyj} for a covariant version of the KS double copy where no particular time coordinate is chosen.}
\begin{equation}
J^{\mu}=2\left[?T^{\mu}_0?-\frac{1}{d-2}?\delta^{\mu}_{0}?T\right]. 
\end{equation}
The time component of (\ref{YM}) is
\begin{equation}
	-\p^2\phi=-\vec{\nabla}^2\phi=gJ^0.\label{pois}
\end{equation}
Due to the linearization of the Ricci tensor for the metrics of the KS form, one obtains the linearized equations of YM theory and biadjoint scalar theory, namely, Maxwell's (\ref{YM}) and Poisson's equations (\ref{pois}). While the gauge field $A_\m$ is called the single copy of the KS graviton $h_{\m\n}$, the scalar field $\phi$ is interpreted as the zeroth copy of the gauge field $A_\m$. 

While the construction was also extended to metrics with multi KS forms \cite{Luna:2015paa}, time dependence \cite{Luna:2016due} and  backgrounds more general than Minkowski \cite{Carrillo-Gonzalez:2017iyj}, the only known way to study metrics with no KS form is to employ perturbation theory \cite{Neill:2013wsa,Luna:2016hge,Luna:2017dtq,Goldberger:2016iau,Goldberger:2017frp,Goldberger:2017vcg,Goldberger:2017ogt,Bern:2019nnu,Chester:2017vcz,Shen:2018ebu,Plefka:2018dpa}. Other developments in the classical double copy include the relation between the sources in the gravity and the gauge theory sides \cite{Ridgway:2015fdl}, and nonperturbative  \cite{Berman:2018hwd,White:2016jzc,DeSmet:2017rve,Bahjat-Abbas:2020cyb} and global \cite{Alfonsi:2020lub}  aspects. However, the restrictive nature of the KS double copy seems to be an obstacle to a better understanding. In \cite{CarrilloGonzalez:2019gof}, the authors used 3D physics as a testing ground for the classical double copy since, at first sight, it is not obvious how it works. General relativity in 3D has no propogating degrees of freedom and therefore the first question that needs to be answered is how the degree of freedom of the photon, which is one in 3D, is matched in the gravity side. The second question is related to the nature of Newtonian and Coulomb potentials in 3D. An application of Gauss' law to a point particle of charge $Q$ and mass $M$ suggests a logarithmic form for both as follows
\begin{equation}
\oint \mathbf{E}\boldsymbol{\cdot} \text{d}\mathbf{A}\propto Q, \qquad E\propto\frac{1}{r}, \qquad \phi \propto \log r,\label{coulomb}
\end{equation}
\begin{equation}
	\oint \mathbf{g}\boldsymbol{\cdot} \text{d}\mathbf{A}\propto M, \qquad g\propto\frac{1}{r}, \qquad \Phi \propto \log r.\label{newton}
\end{equation}
Whereas, in 3D, the Coulomb potential given in (\ref{coulomb}) is a consequence of Maxwell's equations, general relativity has no Newtonian limit giving rise to (\ref{newton}). In taking the Newtonian limit, one considers a weak deviation from the flat space ($g_{\m\n}=\e_{\m\n}+h_{\m\n}$ with $\left|h_{\mu \nu}\right| \ll 1$) and a stationary weak source ($?T^0_0?=-\r$, $?T^i_0?=0=?T^i_j?$) and the geodesic equation reduces to
\begin{equation}
	\frac{\mathrm{d}^{2} \boldsymbol{x}}{\mathrm{d} t^{2}}=\frac{1}{2} \vec{\nabla} h_{00}.
\end{equation}
Using 
\begin{equation}
		\frac{\mathrm{d}^{2} \boldsymbol{x}}{\mathrm{d} t^{2}}=\vec{g}=-\vec{\nabla}\Phi,
\end{equation}
one finds the Newtonian potential as
\begin{equation}
	\Phi=-\frac{1}{2}h_{00}\label{newt}
\end{equation}
However, when $d=3$, the 00 component of Einstein equations becomes
\begin{equation}
	\vec{\nabla}^2\Phi=0,
\end{equation}
resulting in a trivial Newtonian potential $\Phi=0$. If the double copy construction is possible, this problem should be automatically solved since using the 00 component of the KS graviton $h_{00}=\k\phi$ in (\ref{newt}) yields a nontrivial Newtonian potential as
\begin{equation}
	\Phi = -\frac{\k}{2}\phi
\end{equation}
When one starts from the Coulomb solution and achieves a double copy, the logarithmic form of the Coulomb potential is naturally moved into the metric and one obtains a nontrivial solution.

 It turns out that, in 3D, the construction possesses a unique feature with no higher-dimensional counterpart. Although one starts with a vacuum solution in the gauge theory side, one obtains a nonvacuum gravity solution with a nontrivial energy-momentum tensor. In \cite{CarrilloGonzalez:2019gof}, the source was interpreted as a dilaton, which also seems to solve the degree of freedom problem. It is also in agreement with the fact that the double copy of the pure YM theory is gravity coupled to a two-form field and a dilaton, and the absence of the two-form field can be explained by the symmetric nature of the KS ansatz (\ref{met}).

This paper aims to study the construction of \cite{CarrilloGonzalez:2019gof} and further examine the nature of the source by considering a stationary solution, which is a natural generalization of the static solution. Additionally, we introduce a cosmological constant, find solutions in the KS form and present the corresponding gauge theory single copies. The outline of the paper is as follows: In Sec. \ref{sec:static}, we review the main findings of \cite{CarrilloGonzalez:2019gof} and give an alternative way to obtain the static solution. In Sec. \ref{sec:stationary}, we find the stationary version of the solution and show that it is sourced by not a dilaton but a spacelike perfect fluid. Then, we give its gauge theory single copy and discuss some properties of the solution briefly. We end this section by discussing the addition of the cosmological constant. In Sec. \ref{sec:conc}, we conclude with comments on the validity of the classical double copy based on our results.
\section{Static Solution}\label{sec:static}

%\subsection{Banados-Teitelboim-Zanelli Black Hole and Its Gauge Theory Copy}
%We start with the Banados-Teitelboim-Zanelli black hole, which is the most widely studied solution in 3D gravity due to its simplicity

%\subsection{The Static Solution: Free Scalar vs Spacelike Fluid as the Source}

Our starting point is a point charge in 3D Maxwell's theory. In polar coordinates $(t,r,\t)$, the flat space metric takes the form
\begin{equation}
	\e_{\m\n}\dif x^\m \dif x^\n=-\dif t^2+\dif r^2+r^2\dif \t^2,
\end{equation}
and the current vector is given by
\begin{equation}
	J^\m \p_\m=g Q \d^{(2)}(\vec{r})\p_t
\end{equation}
where $Q$ is the charge of the particle. The Coulomb solution is obtained as
\begin{equation}
	A_\m=\phi k_\m, \qquad k_\m \dif x^\m=-\dif t, \qquad \phi=-\frac{g Q}{2\pi}\log r,
\end{equation}
due to the relation $\vec{\nabla}^2 \log r=2\pi\d^{(2)}(\vec{r})$ in two spatial dimensions. In order to obtain a metric in the KS form, we first write the solution in a gauge where the vector $k_\m$ is null as follows
\begin{equation}
A_\m=\phi k_\m, \qquad -k_\m \dif x^\m=\dif t + \dif r, \qquad \phi=-\frac{g Q}{2\pi}\log r,
\end{equation}
Identifying the charge with the black hole mass parameter, $Q\goesto M$, the double copy is given by the metric
\begin{align}
	%$g_{\m\n}= \e_{{\m\n}}+\k \phi k_\m k_\n=...,
	\dif s^2&= \e_{\m\n}\dif x^\m \dif x^\n+\k \,\phi \left(k_\m \dif x^\m\right)^2,\nn\\
	&= -(1+2GM \log r) \dif t^2+(1-2GM \log r) \dif r^2 -4GM \log r\, \dif t \dif r  +r^2\dif \t^2.    \label{metKS}
\end{align}
Note that the vector $k_\m$ is null with respect to both metrics $\e_{{\m\n}}$ and $g_{\m\n}$. The Ricci tensor of the metric (\ref{metKS}) reads
\begin{equation}
	R_{\mu\nu}=\left(\begin{array}{ccc}
	0 & 0 & 0\\
	0 & 0 & 0\\
	0 & 0 & -2GM
	\end{array}\right).
\end{equation}
The $\t\t$ component of the Ricci tensor is nonzero everywhere and the same should be true for the energy-momentum tensor $T_{\m\n}$, implying a nonlocal source. The approach of \cite{CarrilloGonzalez:2019gof} is to consider the coupling of gravity to a free scalar field as
\begin{equation}
S=\int \dif^{3}x\sqrt{-g}\left[\frac{\eps_{1}}{\kappa^{2}}R-\frac{\eps_2}{2}\left(\p \vf\right)^2\right],\label{act}
\end{equation}
where $\eps_i=\pm 1$ ($i=1,2$) control the sign of the kinetic terms and take a negative value for a ghost graviton or a dilaton. The trace-reversed field equations which follow from the action (\ref{act}) are
\begin{equation}
	R_{\m\n}=\frac{\eps}{2} \p_\m \vf \p_\n \vf,\label{phieq}
\end{equation}
where $\eps=\eps_{1} \eps_2$. The metric in (\ref{metKS}) is a solution if $\eps=-1$ and the gradient of the field is 
\begin{equation}
	\p_\m \vf=\sqrt{\frac{M}{2\pi}}(0,0,1),
\end{equation}
which implies that the dilaton is linear in the azimuthal angle. The matter field equation is also satisfied as
\begin{equation}
	\Box \vf \equiv \nabla_\m \nabla^\m \vf = 0.\label{scalar}
\end{equation}
The analysis of \cite{CarrilloGonzalez:2019gof} proceeds by taking $\eps_1=+1$ and $\eps_2=-1$, i.e., the dilaton should be a ghost to support the metric given in (\ref{metKS}). It was also shown that, in a proper generalized gauge, the part of the Lagrangian which is quadratic in the fields can be put in a form where the graviton and the dilaton kinetic terms have nonghost signs, exhibiting a parallelism with the double copy construction in scattering amplitudes. The existence of the scalar hair was attributed to two facts: the scalar field is a ghost and it does not respect the symmetries of the spacetime, i.e., $\p_\m\vf\neq0$. However, the same solution can be obtained by taking $\eps_1 = -1$ and $\eps_2 = +1$, and therefore, the former does not play a role here (see the Appendix for a discussion of the no-hair theorem for free scalar fields).

This choice of the signs has the advantage that the dilaton is not a ghost any more and the ``wrong" sign for the graviton kinetic term has no physical importance since it does not propagate any dynamical degree of freedom. This is, indeed, an approach which is employed to preserve the unitarity of modified gravity theories such as topologically massive gravity \cite{Deser:1981wh,Deser:1982vy} and new massive gravity \cite{Bergshoeff:2009hq}. In this case, the quadratic part of the Lagrangian contains no dynamical ghost, and therefore there is no need for performing a generalized gauge transformation. Hence, one might speculate that it might have also interesting consequences for the double copy in 3D scattering amplitudes, which, we believe, deserves further study.

Motivated by this possibility of obtaining the solution with different sign choices, one might also ask whether there is any other freedom in the construction of the solution, which is answered in \cite{CarrilloGonzalez:2019gof} to a certain extent. It was shown that,  while it is not possible to see the source as a timelike fluid (perfect or viscous), the solution can also be obtained by coupling to a spacelike perfect fluid, whose energy-momentum tensor reads
\begin{equation}
	T_{\m\n}=\left(\r+P\right)u_\m u_\n+P g_{\m\n},\qquad u^2=+1,
\end{equation}
where $u^\m$ is the velocity of the fluid. The Einstein equations in this case become
\begin{equation}
	R_{\m\n}=\frac{\k^2}{2}\left[\left(\r+P\right)u_\mu u_\n - \left(\r+3P\right) g_{\m\n}\right].\label{fluideq}
\end{equation}
Comparing this with (\ref{phieq}) and removing the metric term in (\ref{fluideq}) by choosing $\r=-3P$ gives
\begin{equation}
	R_{\m\n}=-\k^2 P\, u_\m u_\n.\label{fluideq2}
\end{equation}
Therefore, it yields the same solution if the pressure is chosen to be the norm of the gradient of the field $\vf$ as
\begin{equation}
	P=\frac{1}{2}(\p\vf)^2=\frac{M}{4\pi r^2}=-\frac{1}{3}\r
\end{equation}
and the fluid velocity is given by
\begin{equation}
 u_\m=\left(0,0,r\right).
\end{equation}
This alternative reflects the correspondence between scalar fields and perfect fluids \cite{Faraoni:2012hn,Semiz:2012zz,Faraoni:2018fil}. However, as we will show in the next section by studying a stationary solution in the KS form, the correspondence does not always hold and one is forced to choose the spacelike fluid interpretation.

\section{Stationary Solution}\label{sec:stationary}
In this section, we will put different interpretations of the source on a test by studying a more nontrivial solution. To introduce rotation, we write the flat metric in spheroidal coordinates $(t,r,\t)$ as
\begin{equation}
\e_{\m\n}\dif x^\m \dif x^\n=-\dif t^2+\frac{r^2}{r^2+a^2}\dif r^2+(r^2+a^2)\dif \t^2,
\end{equation}
where $a$ will be the rotation parameter. The null vector $k_\m$ is parametrized as
\begin{equation}
	-k_\m \dif x^\m=\dif t + \frac{r^2}{r^2+a^2} \dif r + a \dif \t.
\end{equation}
For a metric in KS form with $\phi=\phi(r)$, the metric becomes
\begin{eqnarray}
ds^{2} &=& \e_{\m\n}\dif x^\m \dif x^\n+\k \,\phi \left(k_\m \dif x^\m\right)^2,\nn\\
& = & -(1-\k\phi(r))\dif t^{2}+\frac{r^{2}\left[a^{2}+r^{2}(1+\k\phi(r))\right]}{\left(a^{2}+r^{2}\right)^{2}}\dif r^{2}+\left[a^{2}(1+\k\phi(r))+r^{2}\right]\dif\theta^{2}\nonumber \\
&  & +\frac{2\k ar^{2}\phi(r)}{a^{2}+r^{2}}\,\dif r\dif \theta+\frac{2\k r^{2}\phi(r)}{a^{2}+r^{2}}\,\dif r\dif t+2\k a\phi(r)\,\dif t\dif\theta,\label{metinKS}
\end{eqnarray}
and  the independent components of the Ricci tensor read
\begin{align}
	R_{00}&=\k\frac{\left(a^{2}-r^{2}+\k r^{2}\phi(r)\right)\phi'(r)+r\left(-a^{2}-r^{2}+\k r^{2}\phi(r)\right)\phi''(r)}{2r^{3}}, \nn\\
	R_{11}&=\k \frac{r\left(a^{2}+r^{2}+\k r^{2}\phi(r)\right)\left(\phi'(r)+r\phi''(r)\right)}{2\left(a^{2}+r^{2}\right)^{2}}, \nn\\
	R_{22}&=\k \frac{\left(a^{4}+3r^{2}a^{2}+\k r^{2}\phi(r)a^{2}+2r^{4}\right)\phi'(r)-a^{2}r\left(a^{2}+r^{2}-\k r^{2}\phi(r)\right)\phi''(r)}{2r^{3}},\nn\\
	R_{01}&=\k^2\frac{r\phi(r)\left(\phi'(r)+r\phi''(r)\right)}{2\left(a^{2}+r^{2}\right)}, \nn\\
	R_{02}&=\k \frac{a\left(\left(a^{2}+r^{2}+\k r^{2}\phi(r)\right)\phi'(r)+r\left(-a^{2}-r^{2}+\k r^{2}\phi(r)\right)\phi''(r)\right)}{2r^{3}},\nn\\
	R_{12}&=\k^2\frac{ar\phi(r)\left(\phi'(r)+r\phi''(r)\right)}{2\left(a^{2}+r^{2}\right)},
\end{align}
where primes denote the derivative with respect to $r$. The trace-reversed Einstein equations when the source is the dilaton (\ref{phieq}) or the spacelike fluid (\ref{fluideq2}) are of the form $R_{\m\n}\propto V_\m V_\n$, where $V_\m$ is a three-vector and one can use this to constrain the function $\phi(r)$. From, for example, $(R_{01})^2=R_{00}R_{11}$, it is easy to see that the only consistent solution takes the form $\phi(r)\propto \log r$ and the proportionality constant is determined by requiring one to get the static solution as $a\goesto 0$, which yields
\begin{equation}
	 \phi(r)=-\frac{\k M}{4 \pi} \log r.\label{phisol}
\end{equation}
Using this in (\ref{phieq}) with again $\eps=-1$, one sees that the gradient of the scalar $\vf$ should be given by
\begin{equation}
\partial_{\mu}\varphi=\sqrt{\frac{M}{2\pi}}\left(\frac{a}{r^{2}},0,\frac{a^{2}+r^{2}}{r^{2}}\right).
\end{equation}
When the rotation is turned on, $a\neq0$, this introduces an $r$ dependence in the $t$  and $\t$ components, which conflicts with the fact that the $r$ component is zero, hence no $r$ dependence. Therefore, unlike the case for the static metric, there is no consistent solution for the function $\vf$.

However, one can still use the spacelike fluid as the source and the metric with the scalar $\phi$  given in (\ref{phisol}), is a solution of Einstein equations (\ref{fluideq2}) when 
\begin{equation}
		P=\frac{1}{2}(\p\vf)^2=\frac{M}{4\pi r^2}=-\frac{1}{3}\r,\qquad u_\m=\left(\frac{a}{r},0,\frac{a^2+r^2}{r}\right).\label{fluid}
\end{equation}
 Therefore, the stationary solution cannot be sourced by a dilaton and the spacelike fluid becomes compulsory to obtain a stationary solution in the KS form.

The gauge theory single copy can easily be obtained as
\begin{equation}
	A_\m\dif x^\m=\phi k_\m\dif x^\m= \frac{g Q}{2\pi}\log r\left(\dif t + \frac{r^2}{r^2+a^2} \dif r + a \dif \t\right),\label{gauge1}
\end{equation}
which is a solution of Maxwell's equations (\ref{YM}) with the current vector
\begin{equation}
	J^\m = \r v^\m, \qquad \r=\frac{Qa^2}{\pi r^4}, \qquad v^\m=\left(1,0,-\frac{1}{a}\right),\label{gauge2}
\end{equation}
which describes a rotating nonlocal charge distribution with angular velocity $\omega=-\sfrac{1}{a}$ with respect to the origin. Checking the nonzero components of the field strength tensor,
\begin{equation}
	F_{rt}=\frac{g Q}{2\pi r},\qquad F_{r\t}=\frac{a g Q}{2\pi r},
\end{equation}
one sees that the magnetic field is created due to the rotation in the gravity side.

In order to see some main properties of the metric (\ref{metinKS}) with the scalar $\phi$ given in (\ref{phisol}), it is useful to write it down in Boyer-Lindquist coordinates, which is achieved by the transformations \cite{Myers:1986un}
\begin{align}
	\dif\theta&\mapsto \dif\Theta+h_{1}\dif r,\nn\\
	\dif t&\mapsto \dif T+h_{2}\dif r,
\end{align}
where
\begin{align}
	h_{1}&=-\frac{\k ar^{2}\phi(r)}{\left(a^{2}+r^{2}\right)\left(a^{2}-\k r^{2}\phi(r)+r^{2}\right)},\nn\\
	h_{2}&=\frac{\k r^{2}\phi(r)}{a^{2}-\k r^{2}\phi(r)+r^{2}}.
\end{align}
In these coordinates, the metric is given by
\begin{equation}
\dif s^{2}=-(1-\k\phi(r))\dif T^{2}+\frac{r^{2}\dif r^{2}}{a^{2}-\k r^{2}\phi(r)+r^{2}}+\left(r^{2}+a^{2}(1+\k\phi(r))\right)\dif\Theta^{2}+2\k a\phi(r)\,\dif\Theta \dif T\label{BLgeneric}
\end{equation}
When the explicit form of the scalar $\phi(r)$ given in (\ref{phisol}) is used, it becomes
\begin{align}
\dif s^{2}=&-(1+2 G M \log r)\dif T^{2}+\frac{r^{2}\dif r^{2}}{a^{2}+r^2\left(1+2 G M  \log r\right)}+\left(r^{2}+a^{2}(1-2 G M \log r)\right)\dif\Theta^{2}\nn\\
&-4a G M \log r \,\dif\Theta \dif T.
\end{align}
From the curvature invariants
\begin{align}
	R^{\mu\nu}R_{\mu\nu}&=\frac{4G^{2}M^{2}}{r^{4}},\nn\\
	R^{\mu\nu\sigma\rho}R_{\mu\nu\sigma\rho}&=4R^{\mu\nu}R_{\mu\nu}-R^{2}=\frac{12G^{2}M^{2}}{r^{4}},\label{invs}
\end{align}
one sees that the metric has a real singularity at $r=0$. For appropriately chosen parameters, it has an event horizon enclosed by an \textit{ergocircle}, which might be thought of a 3D analog of the Kerr black hole sourced by a spacelike fluid. The metric asymptotically takes the form
\begin{equation}
	\left.\dif s^{2}\right|_{r\goesto \infty}=-2 G M \log r\,\dif T^{2}+\frac{\dif r^{2}}{2 G M  \log r}+r^{2}\dif\Theta^{2},
\end{equation}
and therefore, it is not asymptotically flat. However, it is asymptotically locally flat\footnote{To our knowledge, the black hole solution discovered in \cite{Barnich:2015dvt}  is the only other known solution of this type in 3D.} as can be seen by the vanishing of the curvature invariants (\ref{invs}) as $r\goesto\infty$.

When a cosmological constant is introduced, the trace-reversed Einstein equations become
\begin{equation}
R_{\m\n}-2\L g_{\m\n}=-\k^2 P\, u_\m u_\n.\label{fluidlambda}
\end{equation}
This time, the metric given in (\ref{metinKS}) is a solution when 
\begin{equation}
	 \phi(r)=-\frac{\k M}{4 \pi} \log r+\frac{\L}{\k} r^2,
\end{equation}
with the fluid properties given in (\ref{fluid}). The gauge theory single copy is given by
\begin{equation}
A_\m\dif x^\m=\phi k_\m\dif x^\m= \left(\frac{g Q}{2\pi}\log r-\frac{\L}{2g}r^2\right)\left(\dif t + \frac{r^2}{r^2+a^2} \dif r + a \dif \t\right),
\end{equation}
and the nonzero components of the field strength tensor read
\begin{equation}
F_{rt}=\frac{g Q}{2\pi r}-\frac{\L r}{g},\qquad F_{r\t}=a\left(\frac{ g Q}{2\pi r}-\frac{\L r}{g}\right),
\end{equation}
with the magnetic field again created by the rotation. Maxwell's equations are now given by
\begin{equation}
	\p_\n F^{\n\m}=g\left(J^\m_{(\L=0)}+\bar{J}^\m\right),
\end{equation}
where the first current vector $J^\m_{(\L=0)}$ describes the source in the absence of the cosmological constant and is given by (\ref{gauge2}). The second current vector describes a constant charge density filling all space as follows
\begin{equation}
\bar{J}^\m = \r_0 \bar{v}^\m, \qquad \r_0=\frac{2\L}{g^2}, \qquad \bar{v}^\m=\left(1,0,0\right),
\end{equation}
which is the expected effect of adding a cosmological constant in the gravity side.

When written in Boyer-Lindsquit coordinates using (\ref{BLgeneric}), the metric becomes
\begin{align}
ds^{2}=&-(1+2GM\log r -\Lambda r^{2})\dif T^{2}+\frac{r^{2}}{a^{2}+r^2\left(1+2 G M  \log r-\L r^2\right)}\dif r^{2}\nn\\&+\left(r^{2}+a^{2}(1-2GM \log r +\L r^2)\right)\dif\Theta^{2}
+2a(-2GM \log r +\L r^2)\,\dif\Theta \dif T.\label{metlambda}
\end{align}
From the curvature invariants
\begin{align}
	R^{\mu\nu}R_{\mu\nu}&=\frac{4G^{2}M^{2}}{r^{4}}-\frac{8G\Lambda M}{r^{2}}+12\Lambda^{2},\nn\\
	R^{\mu\nu\sigma\rho}R_{\mu\nu\sigma\rho}&=4R^{\mu\nu}R_{\mu\nu}-R^{2}=\frac{12G^{2}M^{2}}{r^{4}}-\frac{8G\Lambda M}{r^{2}}+12\Lambda^{2},
\end{align}
the singularity at $r=0$ is again apparent and an event horizon enclosed by an ergocircle can be identified by a certain choice of the parameters. Taking $\L=-\frac{1}{\ell^2}$, the asymptotic form of the metric is
\begin{equation}
\left.\dif s^{2}\right|_{r\goesto \infty}=-\left(\frac{r}{\ell}\right)^2\dif T^{2}+\left(\frac{ \ell}{r}\right)^2\dif r^2+r^{2}\dif\Theta^{2},
\end{equation}
which is the anti-de Sitter (AdS) spacetime with radius $\ell$. Therefore, the metric given in (\ref{metlambda}) serves as an interesting alternative to the well-known BTZ black hole \cite{Banados:1992wn}, which is, of course, more physical since it is a solution of the cosmological Einstein theory with no matter field like our spacelike fluid.
 
\section{Conclusions}\label{sec:conc}
In this paper, we revisited the static solution constructed in \cite{CarrilloGonzalez:2019gof}. Being able to obtain the static solution with different sign choices for the kinetic terms of the graviton and the dilaton, yielding a quadratic Lagrangian with no propagating ghost field, we claim that the study of scattering amplitudes in 3D might offer an interesting insight into the double copy because the change of sign of the graviton kinetic term is problematic in higher dimensions. 

Turning our attention to a stationary version of the solution in the KS form, we showed that it cannot be sourced by a free scalar field and the source should be a spacelike fluid. Even in this form, it presents itself as an interesting example of the classical double copy where the gauge theory source is a nonlocal rotating charge distribution. By introducing a cosmological constant, we obtained a rotating, asymptotically AdS solution whose single copy gauge field describes an electric field and a magnetic field, which is proportional to the rotation parameter in the gravity side, and the effect of the cosmological constant shows itself in the gauge theory as a constant charge distribution filling all space. Based on the expectation from the scattering amplitudes that the double copy should be given by gravity coupled to a dilaton, obtaining a stationary solution sourced by a dilaton or understanding why it is not possible remains an open problem, whose solution might give a better understanding of the classical double copy. 
\begin{acknowledgments}
	M. K. G. is supported by T\"{U}B\.{I}TAK Grant No 118F091.
\end{acknowledgments}
\appendix*
\section{No-Hair Theorem for Free Scalar Fields}
In Sec. \ref{sec:static}, we have seen that the scalar hair can be obtained by a different choice of the scalar kinetic term than \cite{CarrilloGonzalez:2019gof}. Here, we show explicitly why this is possible by reviewing the formulation of the no-hair theorem for free scalar fields \cite{Bekenstein:1972ny,Hawking:1972qk,Bekenstein:1995un} . Any static metric can be written as
\begin{equation}
\dif s^2=-N^2 \dif t^2+h_{i j}\dif x^i \dif x^j,
\end{equation}
where $N=N(x^i)$ and $h_{i j}=h_{i j}(x^i)$. Then, assuming no time dependence, the equation for the free scalar field (\ref{scalar}) becomes
\begin{equation}
\Box \vf(x^i)=\frac{1}{\sqdet}\p_\m\left(\sqdet\, g^{\m\n} \p_\n\vf\right)\label{boxphi}	
\end{equation}
With the help of relations
\begin{align}
\sqdet&=N\sq{h},\nn\\
\frac{1}{\sq{h}}\p_i\left(\sq{h}\,h^{i j}\p_j\vf\right)&=\cD^i \cD_i \vf,
\end{align}
where $\cD_i$ is the covariant derivative with respect to the spatial metric $h_{i j}$, it can be written as
\begin{equation}
\Box\vf(x^i)=\frac{1}{N}\cD^iN\cD_i\vf+\cD^i\cD_i\vf
\end{equation}
Multiplying by $N\vf$ and integrating over the spatial region $\S$ between the event horizon and infinity yields
\begin{equation}
0=\int_\S \dif ^2x\sqh\, \left[\vf \cD^iN\cD_i\vf+N\vf \cD^i\cD_i\vf\right],
\end{equation}
which, after integrating the first term by parts, becomes
\begin{equation}
0=\oint_{\p\S}\dif S_i\,N\vf \cD^i\vf-\int_\S \dif ^2x\sqh\, N\cD_i\vf \cD^i\vf.\label{IBP}
\end{equation}
The surface integral consists of integrations over the event horizon $\p\S_\text{h}$ and the spatial infinity $\p\S_\text{inf}$ as
\begin{equation}
\oint_{\p\S}\dif S_i\,N\vf \cD^i\vf=\int_{\p\S_\text{h}}\dif S_i\,N\vf \cD^i\vf+\int_{\p\S_\text{inf}}\dif S_i\,N\vf \cD^i\vf
\end{equation}
The first term is zero since the function $N$, by definition, vanishes at the event horizon while the second term is zero provided that the field $\vf$ or its derivative $\cD_i\vf$ vanishes at infinity. With this assumption, (\ref{IBP}) becomes
\begin{equation}
0=\int_\S \dif^2x\sqh\, N\cD_i\vf \cD^i\vf.
\end{equation}
Since the integrand is positive definite, one must have $\cD_i\vf=0$ and therefore $\vf=\text{constant}$ throughout the entire region $\S$. In the usual formulation of the no-hair theorem, one takes $\vf=0$ at the spatial infinity $\p\S_\text{inf}$, which is a reasonable assumption for  physical fields. This implies that the constant should be set to zero, and hence $\vf=0$ , i.e., no scalar can be present in the region $\S$. 

For our analysis, two things are important: First, in the formulation of the no-hair theorem, there is no reference to the sign of the kinetic term of the scalar field in the action since we directly start from its field equation. Therefore, whether it is a ghost or not does not play a role. Second, it becomes possible to obtain a scalar hair because we do not demand $\cD_i\vf=0$ at infinity. Indeed, one has $\cD_i\vf=\text{constant}$ for the static solution.

\end{document}